\documentclass[journal]{IEEEtran}
\usepackage{graphicx}
\usepackage{amsmath}
\usepackage{comment}

\usepackage[font=scriptsize]{caption}
\usepackage{subcaption}
\usepackage{multicol}

\ifCLASSINFOpdf
\else
\fi
\begin{document}
%
\title{Connecting the Unconnected - Sentiment Analysis of Field Survey of Internet Connectivity in Emerging Economies}
%
%
%

\author{Dibakar~Das,
        Barath~S~Narayan,
        Aarna~Bhammar 
        and Jyotsna~Bapat
\thanks{This project is funded by Royal Academy of Engineering, United Kingdom. All authors are affiliated to International Institute of Information Technology, Bangalore (IIITB) India, e-mail: dibakard@ieee.org}
\thanks{Manuscript received TBD; revised TBD.}
}

%
%

\markboth{Journal of \LaTeX\ Class Files,~Vol.~14, No.~8, August~2015}%
{Shell \MakeLowercase{\textit{et al.}}: Bare Demo of IEEEtran.cls for IEEE Journals}
%



\maketitle

\begin{abstract}
Internet has significantly improved the quality of citizens across the world. Though the internet coverage is quite high, 40\% of global population do not have access to broadband internet. This paper presents an analysis of a field survey of population in some areas of Kathmandu, Nepal, an emerging economy. This survey was triggered by intermittent severe congestion of internet in certain areas of the city. People from three different areas were asked about their present experience of internet usage, its impact on their lives and their aspirations for the future. Survey pointed to high speed, low cost, reliable and secure internet as a major aspiration of the respondents. Based on their inputs, this paper presents a sentiment analysis as well as demographic information. Keys insights from this analysis shows that overall sentiment to most queries are positive. The variances of positive sentiments are high whereas those for negative ones are low. Also, some correlations and clusters are observed among the attributes though no dominant component exists in the data.
\end{abstract}


%
\IEEEpeerreviewmaketitle

\section{Introduction}
The internet has significantly transformed human civilization affecting nearly every aspect of life, e.g., communication, education, health care, economy, society, etc. One of the most significant impacts has been on communication. The internet has made it possible to instantly connect with people across the globe. As \cite{cite_rise_network_society} describes the internet has created a "network society," where social, economic and political activities are increasingly mediated through internet.
In terms of education, the internet has democratized access to knowledge. Online learning platforms, open access journals and educational resources have made it easier for individuals to acquire new skills and pursue education without traditional barriers. A study in \cite{cite_digital_scholar} highlights how the internet has enabled the rise of Massive Open Online Courses (MOOCs) transforming education by making it more accessible and flexible to citizens across the world.
The internet has also reshaped the economy, giving rise to the digital economy. Research in \cite{cite_second_machine_age} discusses how the internet, combined with automation and artificial intelligence (AI), has led to significant changes in employment patterns and economic growth.
Socio-culturally, the internet has facilitated the globalization of media allowing rapid spread of ideas and cultural influences turning the world a "global village," with free flow of information flows freely across borders \cite{cite_understanding_media}.

The internet has significantly impacted emerging economies, transforming various sectors like education, health care, business and governance \cite{cite_world_dev_report}. It has provided new opportunities for growth. One of the primary benefits of the internet in emerging economies is its role in bridging the digital divide, improving access to information and services.
In education, the internet has revolutionized learning opportunities in areas where traditional education infrastructure is limited, e.g., MOOCs, enabling people in emerging economies to access high-quality education \cite{cite_moocs_musings}.
In health care, internet-based solutions have played a vital role in improving medical access. Telemedicine and health applications provide remote consultations and access to medical advice particularly in rural or underserved areas \cite{cite_telemedicine_developing_countries}.
Economically, the internet has facilitated the growth of small and medium-sized enterprises (SMEs), which are often the backbone of emerging economies. Online platforms such as e-commerce sites, digital marketing, and mobile payment solutions have opened global markets to local entrepreneurs \cite{cite_mobiles_economic_dev_africa}.

Though, there are several benefits of internet \cite{cite_socio_economic_benefits_of_broadband}, still 40\% of population of the work do not have access to broadband internet \cite{cite_digital_inclusion_world_bank}. Interesting, the coverage of mobile internet has significantly improved the adoption of internet seems to be low probably due to factors such as, affordability, lack of literacy, safety and security and connectivity experience \cite{cite_gsma_report_mobile_internet}. Several studies have been conducted to bridge this divide \cite{cite_bridging_digital_divide}.

Field surveys play a key role in understanding the ground realities prevailing in the society on a particular topic of interest. It provides an insight in the requirements of the society in that subject matter.
A spatio-temporal study analyzed global internet user percentage trends from 1996 to 2017 and found that countries exhibited clustered spatial patterns and a decreasing disparity in internet usage over time and a trend toward more balanced global internet development \cite{cite_spatio_temporal_internet_users}.
Analysis of how internet influences human development across 180 countries from 2010–2017, using mobile broadband and internet bandwidth as key indicators and shows how mobile broadband improves health and education in developing countries, while increased internet bandwidth boosts income in developed countries \cite{cite_mobile_broadband_internet_bw_development}.
A structured review of research on the economic effects of broadband internet, focusing on both wireline and wireless technologies and their availability and adoption is presented and explores impacts on growth, employment and productivity, and highlights research gaps while offering guidance for policymakers on broadband’s role in enhancing social welfare \cite{cite_broadband_internet_economic_impacts}. A study on impact of telecommunications on economic growth finds that the relationship between the two is non-linear with a more significant impact on developing countries, and observes a kink effect where the influence of telecommunications varies depending on the country's development stage \cite{cite_telecom_impact_on_economics}.
A study on internet usage during adolescence has been conducted in \cite{cite_internet_usage_adolescence}.
A research on the internet usage patterns of undergraduate students, focusing on how they use the internet for educational purposes and their overall engagement with information technologies and concludes that internet use can be productive for students and help them to consume and also produce technology content and improve their knowledge \cite{cite_internet_usage_undergrads}. Though, the above research works study extensively in different aspects this paper conducts a field survey with the subjects interviewed based on a questionnaire involving the current experience of internet and their future aspirations.

Kathmandu and its surrounding areas was facing a problem of intermittent internet outages in a day which impacting at different levels to its citizens belonging to different socio-economic strata. In this project, a survey was conducted in the affected areas to understand the reason behind the internet outages. Along with the basic problem, the survey also asked question on internet usages and the benefits of using it as well as their aspirations for the future. The key findings from the survey are as follows. The network deployment does not seem to be major issue at least in some cases. Rather, pricing seems to be the major issue. Since, the cost of cellular data services are more users tend to move to Wi-Fi which is much cheaper. When more users move to Wi-Fi, it congests the corresponding network leaving the cellular network under-utilized. This happens at specific times during the 24-hours such as in the evenings when many users use the Wi-Fi network. Once, the Wi-Fi network is fully congested, frustrated users leave the network and then progressively the network performance improves. Thus, if the cellular data charges are brought down, the congestion of Wi-Fi can be avoided as some users may prefer cellular data. Affordability remains a key factor for high speed internet adaptation. The key aspirations of users is to have high-speed internet and reliable and safe internet services for their progress at a low cost. Areas with inadequate telecommunications infrastructure makes the case even more difficult. Thus, there is a need for huge investments in this sector as well as the challenge of innovative business models to have decent return on investments at the same time meet the affordability constraints of the users. The paper presents deeper insights into these broad observations with demographic information and a large language model (LLM) based sentiment analysis of the queries asked to the users. The sentiment is primarily positive on most queries. However, the variances in positive sentiments are much higher than those of negative sentiments.


Rest of the paper is organized as follows. Section \ref{section_survey_method} provides an overview of the field survey conducted and the sentiment analysis method. Survey insights and sentiment analysis are presented in section \ref{section_results}. Key points of this study is discussed in section \ref{section_discussion}. Section \ref{section_conclusion} concludes this paper along with some future directions of research.


\section{Field Survey Overview}\label{section_survey_method}
The primary trigger for this survey was a news that the people of Kathmandu experience severe network congestion intermittently during the entire 24 hours although the network coverage and bandwidth under normal conditions is not that bad.
To address this issue, a study  was conducted a study to understand the existing situation and the expectation of the users for reliable networks to support their needs. The survey had the following objectives.
\begin{enumerate}
\item	Deep dive into the connectivity issue in Kathmandu
\item	Design intervention based on the research outcome
\item	Socio-technical feasibility study
\end{enumerate}

Three different types of locations were identified for the survey in Kathmandu – a commercial hub with high population density, a low-income settlement and a mid to high income settlement area.
\begin{enumerate}
\item   \emph{Commercial Hub (Asan)}:
        The selected commercial hub is a busy  street bustling with business activities. It has  narrow alleys  with small business establishments. The reason behind the selection of this place for the survey is because of  high population and housing densities, its spatial layout and the high adaptation digital services to run businesses. Users in this area use internet for digital banking, communication and entertainment. Business owners communicate with their clients and suppliers over internet. Some interviewees also spend their leisure time on entertainment activities. They use Facebook to stay in touch with family members overseas, watch movies, listen to music on YouTube, and read news online. The younger generation, in particular, uses the internet to play online games on their mobile devices and engage with platforms like Facebook, YouTube, and Instagram.
        People prefer Wi-Fi over cellular data because the price of the latter is much more than the former. Hence, many users connect to one Wi-Fi router which lead to network congestion. Network congestion was more prominent during the evenings which slows down key applications like mobile banking.

        One of the main challenges in this area is that multiple users often connect to a single Wi-Fi router, causing network congestion. Users complained about connectivity issues due to heavy network traffic, especially in the evenings. Slow buffering during mobile banking was also a concern. Interviewees reported experiencing sudden internet slowdowns, which frustrated them and they tried to turn the Wi-Fi -router on and off, change password or use a cellular data package (which is costly) from their network provider. These slowdowns typically occurred when many people were using Instagram reels, playing games, accessing mobile banking, or uploading pictures to social media. A few interviewees mentioned poor network connectivity in narrow alleyways surrounded by tall buildings, and some also noted a lack of internet access during natural disasters.
\item	\emph{Urban Low-income Area (Bansighat)}:
        This locality is a low income and flood prone region. Hence, they do suffer from economic difficulties.
        Majority of respondents used the internet for entertainment (e.g., movies, series, games, songs). Next important usage of the internet was personal and professional communication.
        Respondents use apps like Facebook, Viber, and WhatsApp to stay connected with their family, friends, and relatives. One user shared that the internet has been a source of support in fighting depression by keeping her connected with loved ones. Additionally, many people rely on the internet for making online payments.
        Major problems faced in this area are slow down and buffering during video streaming in non-working hours like early morning, night and weekends due to higher number of users using internet from home. 
        Cellular data service costs differ depending on the package, and most users find them to be expensive. As a result, Wi-Fi-based broadband connectivity is favored over cellular data. Additionally, the majority of users expressed dissatisfaction with slow internet speeds.
\item	\emph{Urban High-income Area (Kusunti)}:
        Kusunti is a community of well-to-do citizens. Most users turn to the internet for entertainment, such as watching movies, series, playing games, and listening to music during their free time. They also rely on apps like Facebook, Viber, and WhatsApp to stay connected with family, friends, and relatives. Several users noted that they use the internet to help decide when and where to buy products, especially when discounts are involved. A few interviewees mentioned using the internet to learn cooking. Female respondents highlighted using the internet for informal education, while some used it for research in their academic fields. Many interviewees utilized the internet for mobile banking, and some shared that it helped their businesses with online marketing and transactions.
        Slow internet remains a significant issue for many users. Some respondents noted that the internet becomes unreliable and slows down during crucial tasks, causing frustration. It tends to be especially slow during peak hours, and high-definition videos often fail to play smoothly. Additionally, the high cost of internet service was also mentioned as a problem.
        Like in other areas, affordable internet services through Wi-Fi are preferred, as cellular data is costly. All respondents consider high-speed internet essential. They also prioritize the security of their online data and transactions.
\end{enumerate}

It was a small field survey given the availability of resources with number of subjects around 100 from all three areas. However, it pointed out very interesting insights on internet usage and aspirations of the people in a major city in a developing country. The following sections present a analysis of the survey containing the demographic information and sentiment analysis of specific questions asked to the respondents.

\subsection{Sentiment Analysis Model}
The sentiment analysis of the textual responses is performed using Large Language Model (LLM). The LLM described in \cite{cite_llm} is used for this purpose. 
This model was selected as it was trained on the multilingual sentiment dataset and due to its effectiveness in correctly predicting the sentiment. If a text is fed to the model, it returns a positive, negative and neutral sentiment. For example, if the sentence “ I like you” feed to the model, the prediction will be positive and a confidence score is returned. This value is then scaled between -1 and 1. The values less than 0, values greater than 0 and values very close 0 are considered as negative, positive and neutral sentiments respectively.
\section{Results and Discussion}\label{section_results}
These section presents the results of the survey conducted in some areas of Kathmandu, Nepal. The survey was not very big given the limited resources, with around 100 subjects across three areas in Kathmandu. The first few sections discusses the demographic and socio-economic aspects of the survey and the latter sections deals with the sentiment analysis of the survey. 

\subsection{Age distribution}
\begin{figure}[ht]
\centering
\includegraphics[width=\columnwidth]{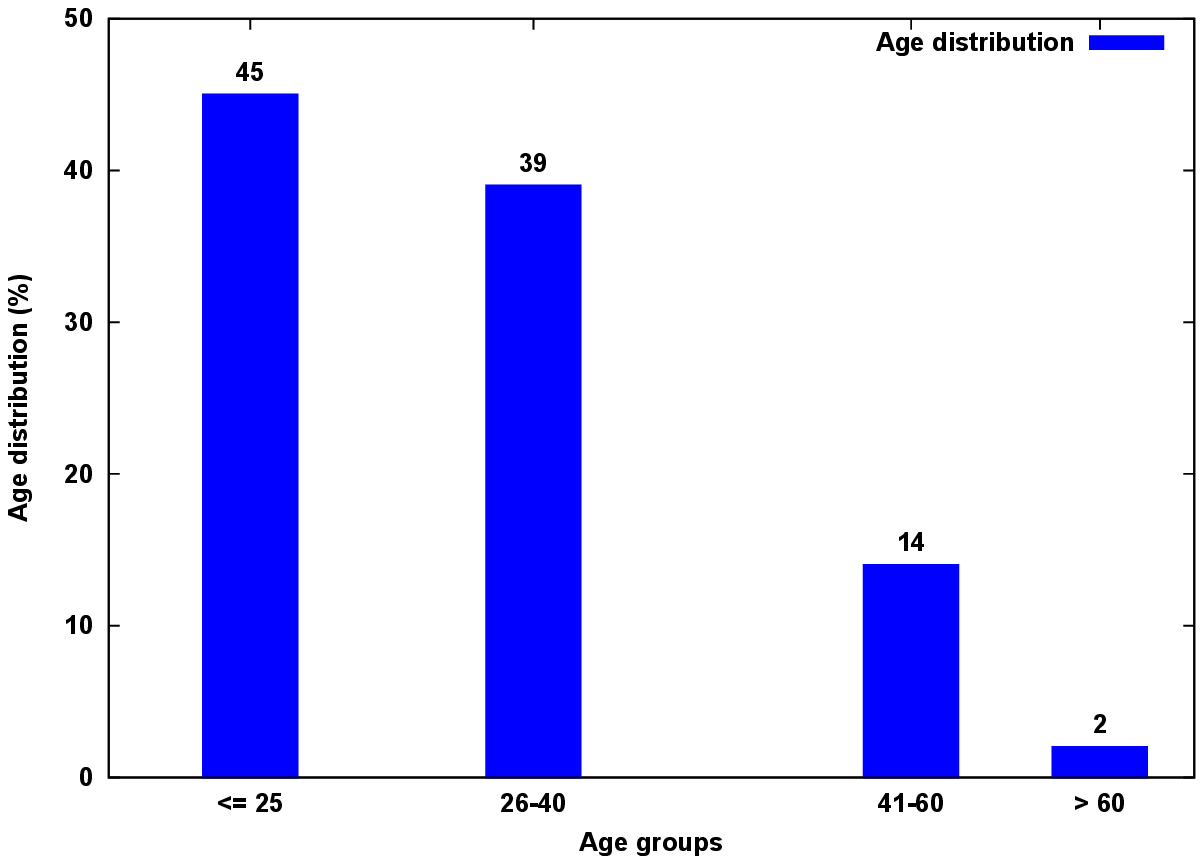}
\caption{Age distribution}
\label{plot_age_distribution_inkscape}
\end{figure}

Fig. \ref{plot_age_distribution_inkscape} shows the age distribution of subjects of the field survey conducted along \emph{y}-axis. The figure shows that the sample is more biased towards young people with $85\%$ of the respondents $\le 40$ years of age. Thus, the survey primarily shows the views of the younger generation.

\subsection{Gender distribution}
\begin{figure}[ht]
\centering
\includegraphics[width=\columnwidth]{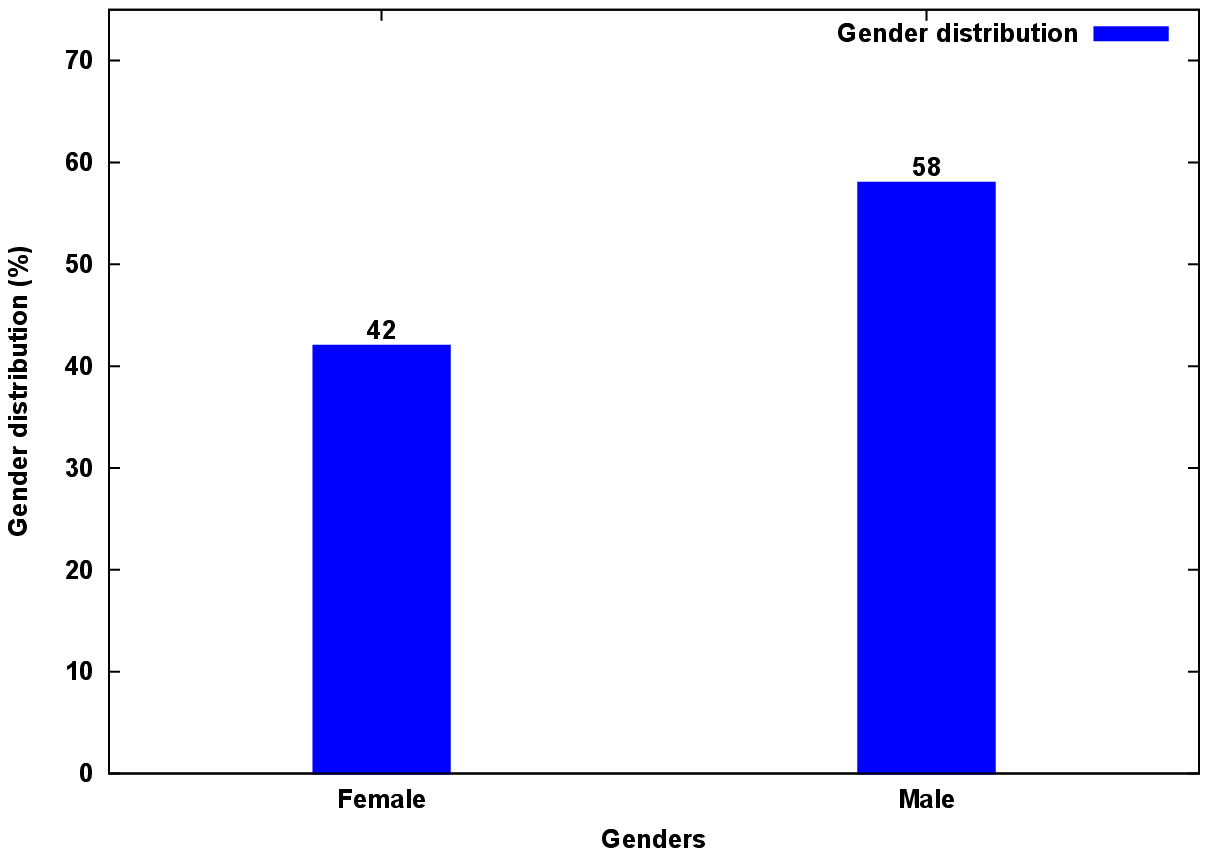}
\caption{Gender distribution}
\label{plot_gender_distribution_inkscape}
\end{figure}

The gender distribution is shown along \emph{y}-axis in Fig. \ref{plot_gender_distribution_inkscape}. The plot shows around $42\%$ of the respondents are female and the rest $58\%$ are males. Female to male ratio is roughly $2:3$.

\subsection{Education distribution}
\begin{figure}[ht]
\centering
\includegraphics[width=\columnwidth]{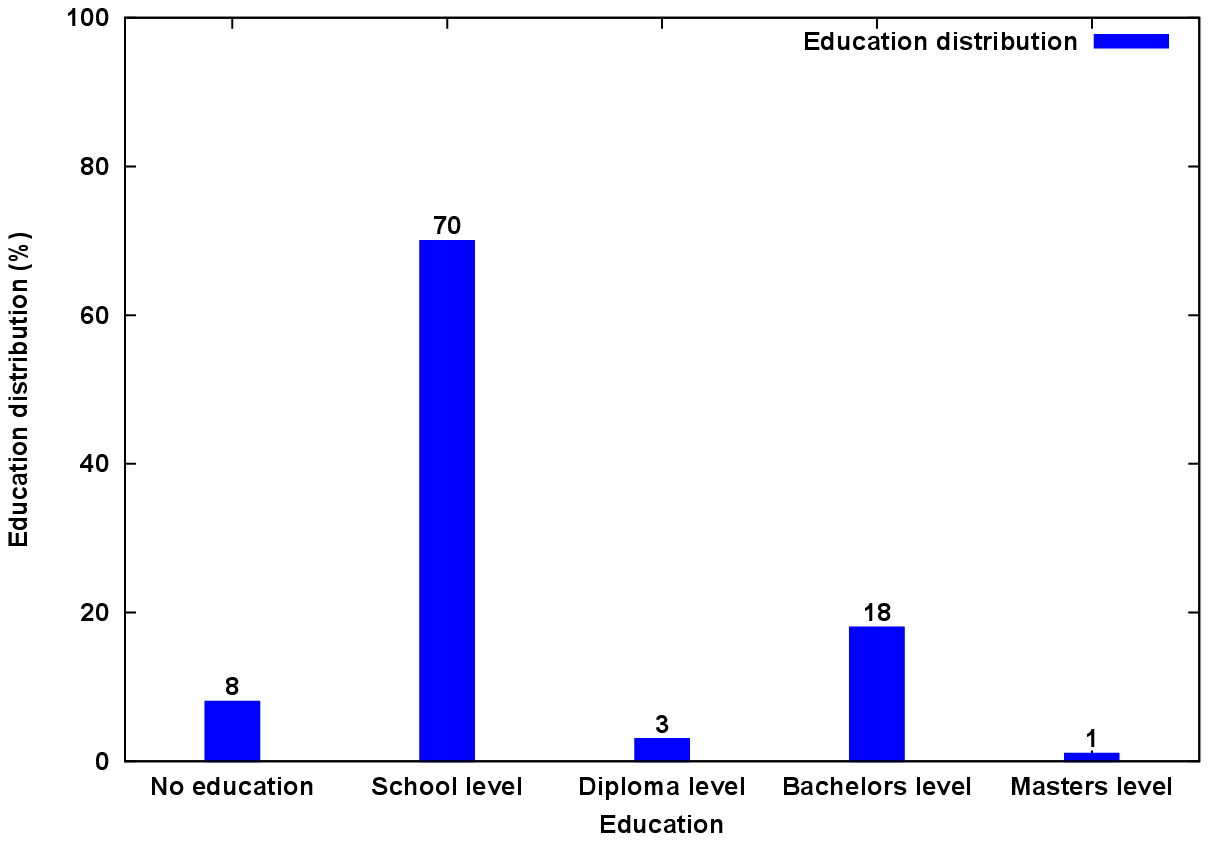}
\caption{Education distribution}
\label{plot_education_distribution_inkscape}
\end{figure}

The education level of the respondents are plotted in Fig. \ref{plot_education_distribution_inkscape} along \emph{y}-axis. It is dominated by people with school level  education ($70\%$) followed by bachelors ($18\%$) and those without any education ($8\%$), and small fractions of diploma holders and masters.

\subsection{Occupation distribution}
\begin{figure}[ht]
\centering
\includegraphics[width=\columnwidth]{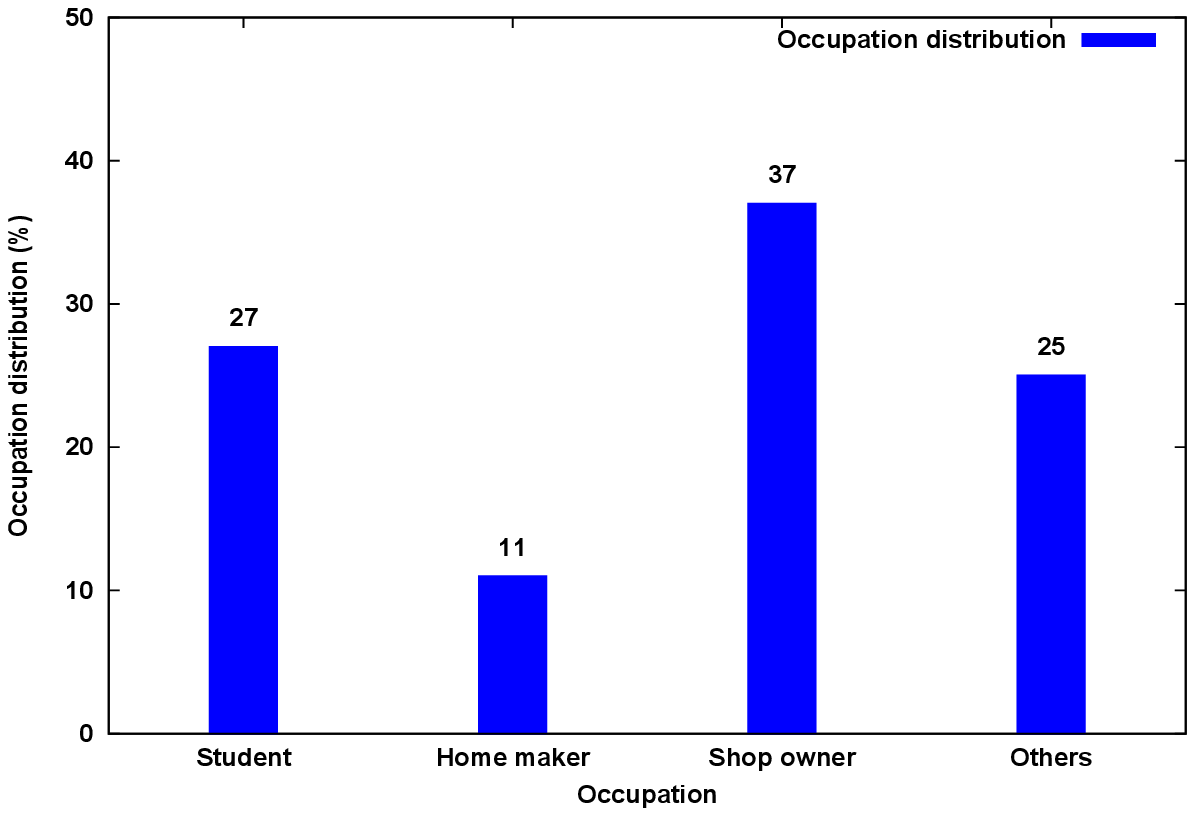}
\caption{Occupation distribution}
\label{plot_occupation_distribution_inkscape}
\end{figure}

Fig. \ref{plot_occupation_distribution_inkscape} shows the occupations of the respondents along \emph{y}-axis. The sample shows groups of shop owners, students, home makers  and a significant number of respondents with heterogenous profession, e.g., working in repair shop, etc.

The above four sections shows the demographic and socio-economic information of the survey. Even though the survey is not very big, it does show a good amount of diversity.
\subsection{User rating of internet experience}
\begin{figure}[ht]
\centering
\includegraphics[width=\columnwidth]{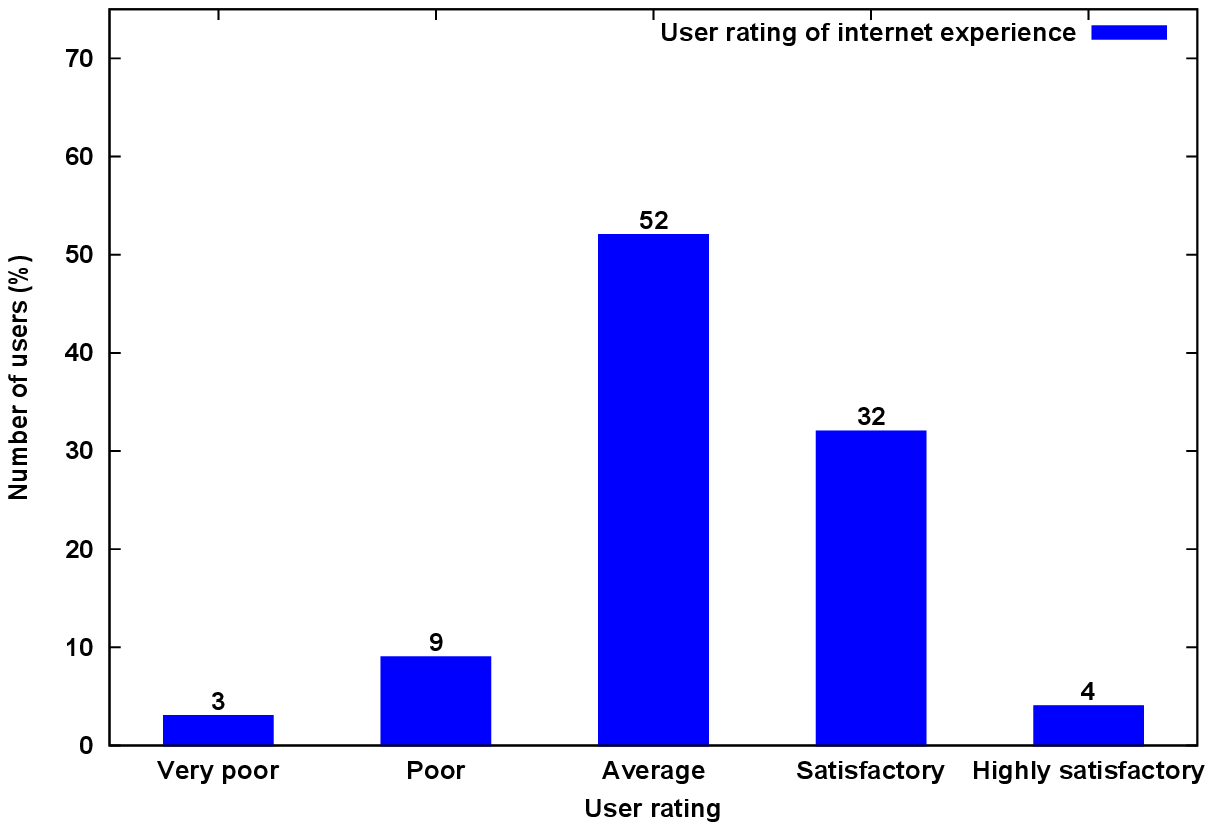}
\caption{User rating of internet experience}
\label{plot_user_rating_inkscape}
\end{figure}

Fig. \ref{plot_user_rating_inkscape} shows the ratings of user satisfaction with the current internet services (along \emph{y}-axis). The 52\% of the users say their the services are average followed by 32\% users who say they are just satisfied. Only 4\% users say they are very satisfied. Rest of the users experience poor services. Thus, there is substantial scope of improvement in user experience of internet to bring 64\% of the respondents to the satisfactory level.

\subsection{Uplink and downlink speeds}
\begin{figure}[ht]
\centering
\includegraphics[width=\columnwidth]{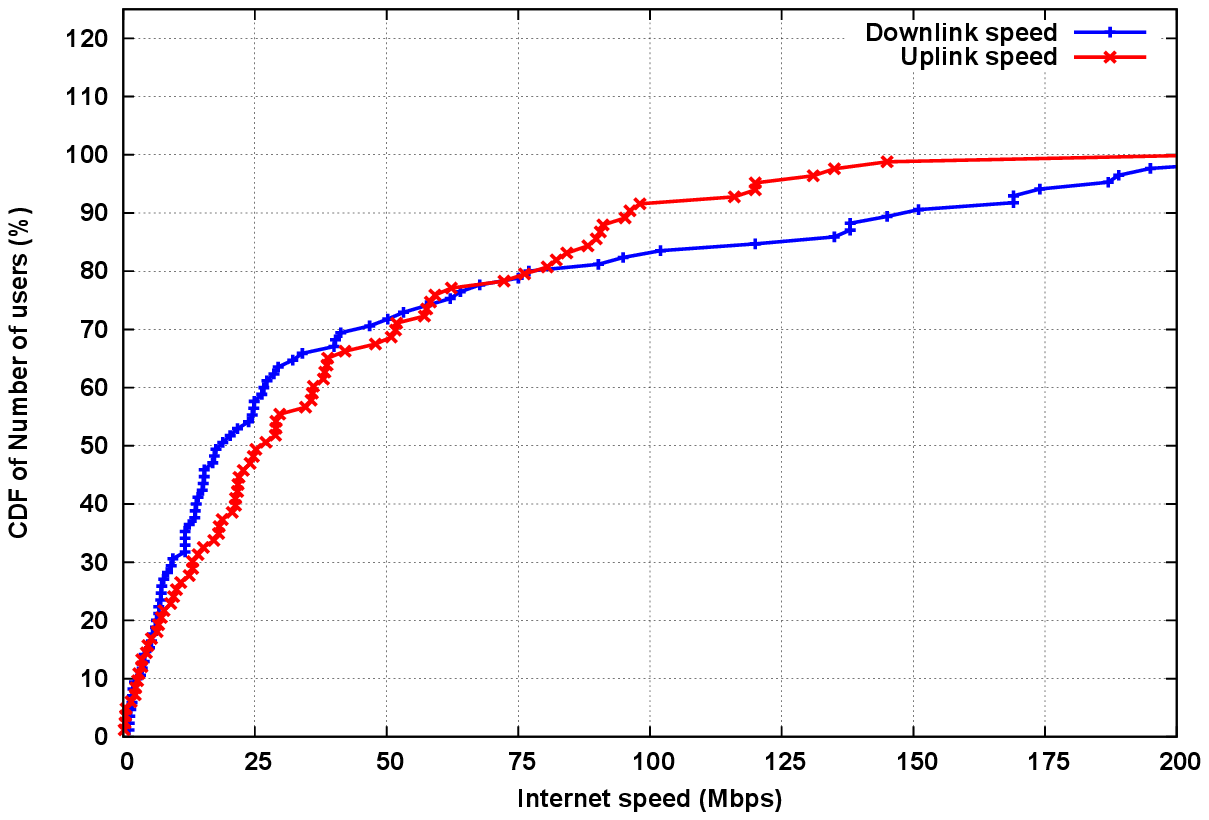}
\caption{CDF of Uplink and downlink speeds}
\label{plot_ul_dl_cdf_percentage_inkscape}
\end{figure}

One of the key metric to evaluate the internet performance is to measure the uplink and downlink speeds. The cummulative distribution of the users (\emph{y}-axis) versus the speed (\emph{x}-axis) is plotted in Fig. \ref{plot_ul_dl_cdf_percentage_inkscape}.  40\% of users are around the 10-12 Mbps mark, 70\% of them experience below 50 Mbps speed and 90\% are below 100 Mbps. Barely, 10\% of users experience close to 100Mbps speed which is important for many professional activities.

Even though the speeds are not that bad for basic services such as browsing and social media the key problem that was observed is intermittent and bursty disruption of internet services during the day. This leads to dissatisfaction among users.

\subsection{Sentiment Analysis}
This section presents the results of sentiment analysis using the multilingual \emph{distillbert} sentiment analysis LLM. The output sentiment values are then scaled between -1 to 1 for negative and positive remarks respectively and 0 is taken as the neutral value. Adding all the values provides the overall sentiment. The mean and variance of positive and negative sentiment values are also evaluated. Number of respondents with positive and negative values provides the fraction of people who provided the corresponding feedbacks.

\subsubsection{Sentiment 1 - How resilient is the network infrastructure to disruptions and disasters? Since 2015 (Earthquake), how has the internet infrastructure changed in your area?}
\begin{figure}[ht]
\centering
\includegraphics[width=\columnwidth]{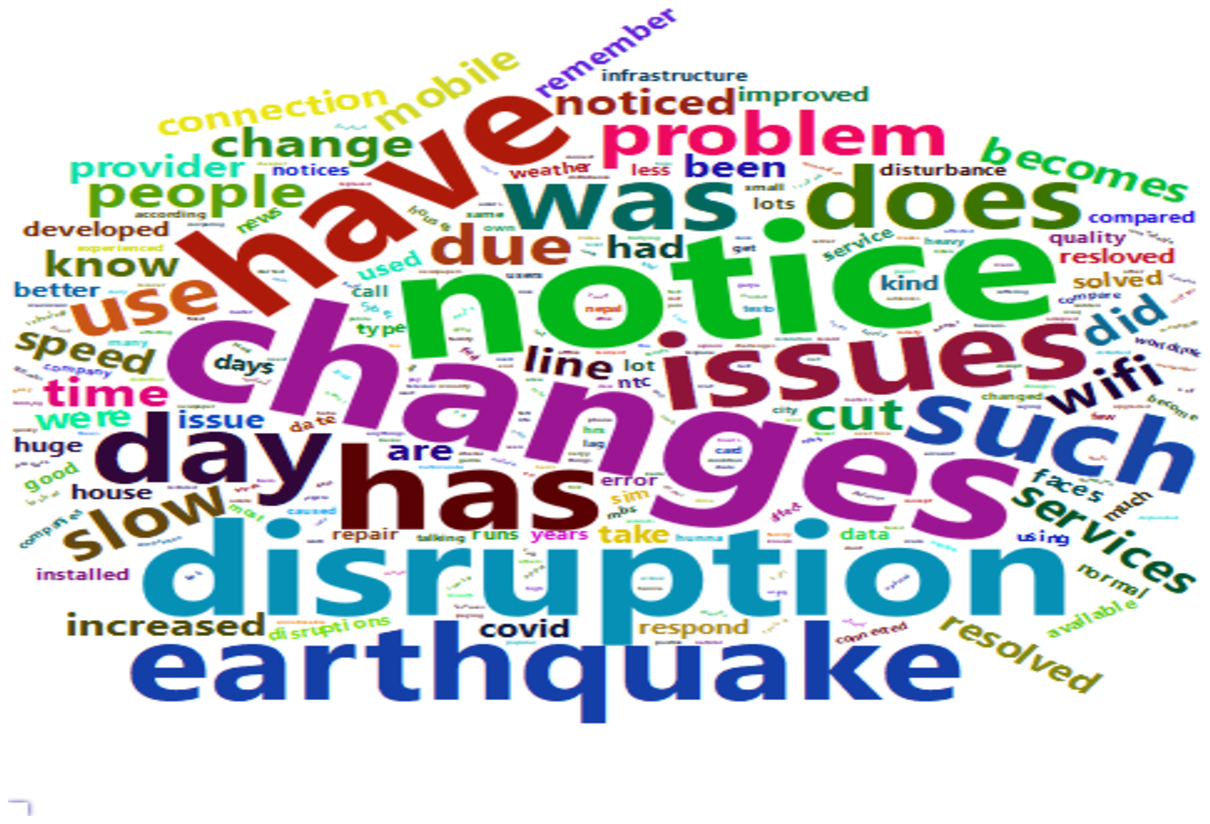}
\caption{Word cloud - Sentiment 1}
\label{plot_sentiment_1_inkscape}
\end{figure}

The above question was asked to the respondents on resilience of the networks especially during disasters. Respondents agree that there are network disruptions during disasters but agree there has been changes and improvements in the services. Lot of them also see that the services disruptions are minor and managed better. The word cloud of the responses is shown in Fig. \ref{plot_sentiment_1_inkscape}.

Significant portion of 56\% say that they do not see any issue either way hence the sentiment is neutral (Table \ref{table_sentiment_1_scores}). 14\% say they strongly feel that the situation has improved and 36\% strongly feel that there are serious disruption of internet during disasters. Hence, overall sentiment is negative. The mean values of positive and negative sentiments are 0.57952 and -0.53072 respectively and both around the value of absolute value 0.5. The variances of positive sentiment is double that of positive sentiment though the values are small. This indicates that the negative sentences more focussed whereas positive sentiments are a bit more blurred. Other reason could be due to the number of subjects with positive sentiment is higher than the negative ones.

The sentiment analysis model seems to take hard (positive or negative) decisions on the text provided. For example, the comment \emph{"Becomes slow during any type of disaster or disruption"} is treated as a strong negative sentiment with value $-0.73102$ whereas \emph{"services is very nice. After technology advances, internet services get better"} is given a strong positive sentiment of value $0.86004$. However, \emph{"Within 1 day it will be resolved"} is treated as neutral value of 0 probably because there are no qualifying words attached to statement.
\begin{table}[ht]
  \caption{Sentiment 1 - Scores}
  \centering
  \begin{tabular}{p{2.5cm}|p{1.5cm}|p{3cm}}
  \hline
  \hline
  \textbf{Parameters} & \textbf{Values} & \textbf{Comments}\\  [0.5ex]
  \hline
  Positive & $14$\% & Subjects who expressed positive sentiment\\
  Negative & $30$\% & Subjects who expressed negative sentiment\\
  Neutral & $56$\% & Subjects who expressed neutral sentiment or did not respond to the query\\
  \hline
  Overall & -7.32651 & Adding all the sentiment values\\
  \hline
  Positive(Mean)	&	0.57952 & Mean  of all positive sentiment values\\
  Negative(Mean)	&	-0.53072 & Mean of all positive sentiment values\\
  Positive(Variance)	&	0.02029 & Variance of all positive sentiment values\\
  Negative(Variance)	&	0.01247 & Variance of all positive sentiment values\\
  \hline
  \end{tabular}
  \label{table_sentiment_1_scores}
\end{table}

\subsubsection{Sentiment 2 - What is the cost of internet services in the area and how does it impact you and your family access to the internet?}
\begin{figure}[ht]
\centering
\includegraphics[width=\columnwidth]{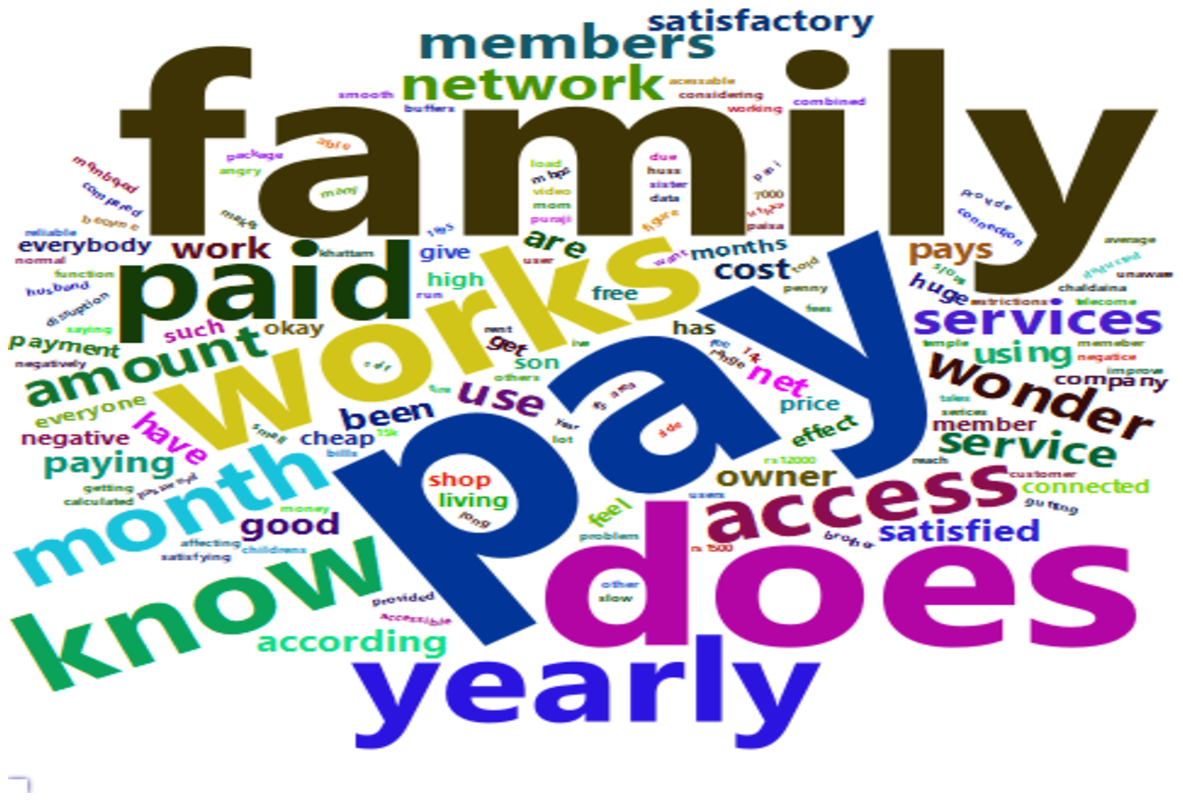}
\caption{Word cloud - Sentiment 2}
\label{plot_sentiment_2_inkscape}
\end{figure}

The above question was asked about the tariffs of internet services. Most people considered the charges as decent when the services are shared by the entire family. Also, some users are provided with free internet by their proprietors. The word cloud of the responses is shown in Fig. \ref{plot_sentiment_2_inkscape}.

The overall sentiment is very positive with a score of $21.27460$ (Table \ref{table_sentiment_2_scores}). Also, $54\%$ of the respondents are satisfied with the internet tariffs. One such positive comment was \emph{"It works well for everyone"}. Roughly $20\%$ of the subjects find the cost of internet high and are unhappy with the services. This was conveyed as one of them as \emph{"The network is slow and I'm not satisfied with the Internet I'm getting compared to the price I pay."}. Among the neutral sentiments of $26\%$, one of them conveyed saying \emph{"I have not calculated"}. The mean values of positive and negative sentiments are $0.60825$ and $-0.54147$ respectively and their variances are in similar ranges with the latter being slightly lower.
\begin{table}[ht]
  \caption{Sentiment 2 - Scores}
  \centering
  \begin{tabular}{p{2.5cm}|p{1.5cm}|p{3cm}}
  \hline
  \hline
  \textbf{Parameters} & \textbf{Values} & \textbf{Comments}\\  [0.5ex]
  \hline
  Positive & $54$\% & Subjects who expressed positive sentiment\\
  Negative & $20$\% & Subjects who expressed negative sentiment\\
  Neutral & $26$\% & Subjects who expressed neutral sentiment or did not respond to the query\\
  \hline
  Overall & 21.27460 & Adding all the sentiment values\\
  \hline
  Positive(Mean)	&	0.60825 & Mean  of all positive sentiment values\\
  Negative(Mean)	&	-0.54147 & Mean of all positive sentiment values\\
  Positive(Variance)	&	0.01528 & Variance of all positive sentiment values\\
  Negative(Variance)	&	0.01219 & Variance of all positive sentiment values\\
  \hline
  \end{tabular}
  \label{table_sentiment_2_scores}
\end{table}
\subsubsection{Sentiment 3 - How do you report internet connectivity issues or provide feedback about the quality of their services?}
\begin{figure}[ht]
\centering
\includegraphics[width=\columnwidth]{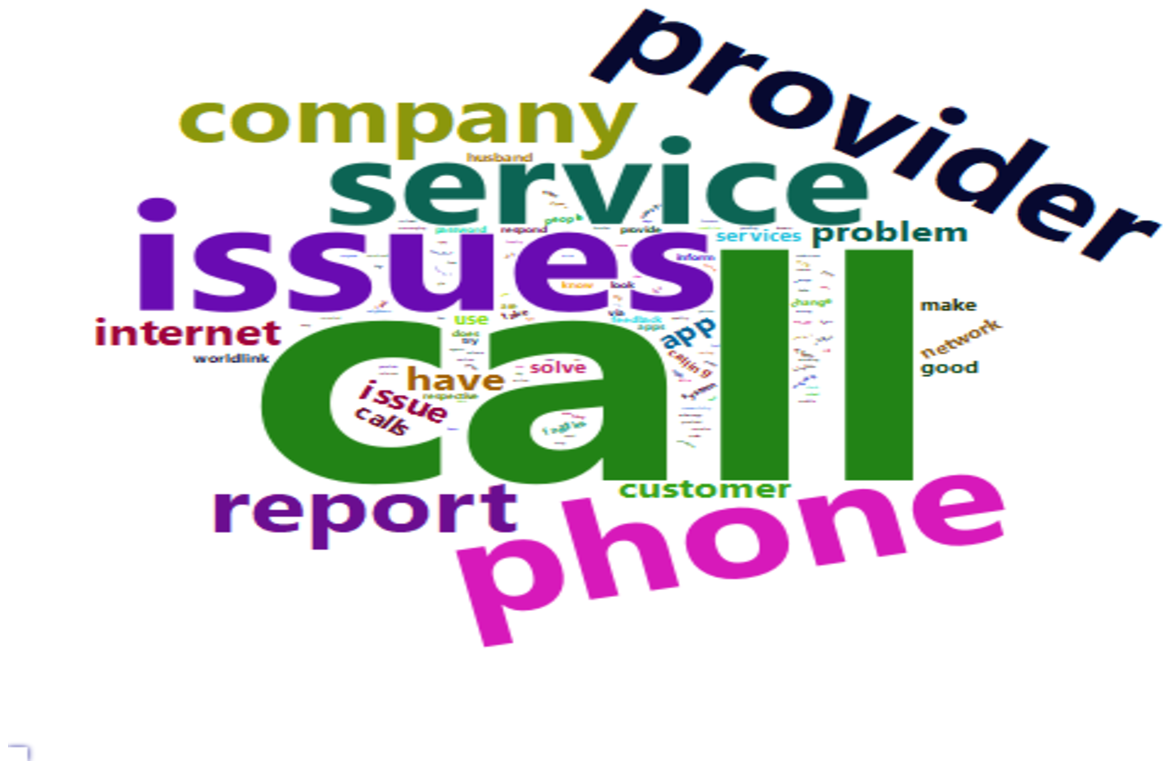}
\caption{Word cloud - Sentiment 3}
\label{plot_sentiment_3_inkscape}
\end{figure}

For the above question, the subjects are again satisfied with the services provided by the internet operators. Though, majority of them agree that there are disruptions in internet services but they normally makes phone calls to the service providers to get the issues resolved. This is also evident from the word cloud in Fig. \ref{plot_sentiment_3_inkscape}.

For this query also the overall sentiment is positive with the overall sentiment of $25.76114$. An overwhelming, $71\%$ are satisfied with the service of the internet providers as evident from one of the responses \emph{"Through phone call. The service is very good. The person from that company comes to the door and look out for the problem and solve it in a minute."}. Around $15\%$ are not particularly happy with the services as one of them responds with \emph{"I contact via chat app but mostly through phone calls which remains pending for 3 to 4 days. I don't really like their customer service."}. Another 15\% says they haven't faced any issues so unable to rate the services, e.g., \emph{"Phone call probably but there has been no complaints yet so I haven't made any phone calls."}. The mean values respectively for positive and negative sentiments are 0.48659 and -0.45388 and their variances are low.
\begin{table}[ht]
  \caption{Sentiment 3 - Scores}
  \centering
  \begin{tabular}{p{2.5cm}|p{1.5cm}|p{3cm}}
  \hline
  \hline
  \textbf{Parameters} & \textbf{Values} & \textbf{Comments}\\  [0.5ex]
  \hline
  Positive & $71$\% & Subjects who expressed positive sentiment\\
  Negative & $15$\% & Subjects who expressed negative sentiment\\
  Neutral & $14$\% & Subjects who expressed neutral sentiment or did not respond to the query\\
  \hline
  Overall & 25.76114 & Adding all the sentiment values\\
  \hline
  Positive(Mean)	&	0.48659 & Mean  of all positive sentiment values\\
  Negative(Mean)	&	-0.45388 & Mean of all positive sentiment values\\
  Positive(Variance)	&	0.00913 & Variance of all positive sentiment values\\
  Negative(Variance)	&	0.00587 & Variance of all positive sentiment values\\
  \hline
  \end{tabular}
  \label{table_sentiment_3_scores}
\end{table}
\subsubsection{Sentiment 4 - Any government policies you wish were in place to better the internet connectivity and infrastructure of Kathmandu?}
\begin{figure}[ht]
\centering
\includegraphics[width=\columnwidth]{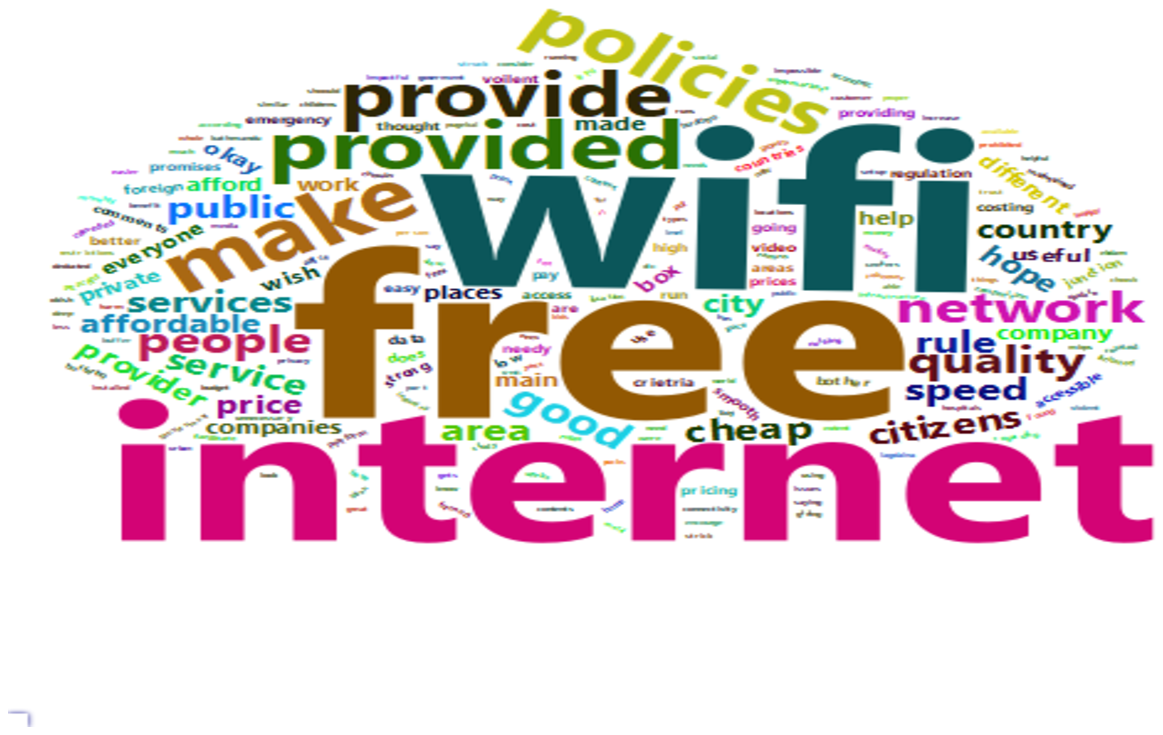}
\caption{Word cloud - Sentiment 4}
\label{plot_sentiment_4_inkscape}
\end{figure}

When the above question was asked to the subjects all agreed that internet services are satisfactory with intermittent congestions. However, the aspirations are very high. Many of the respondents suggest that internet should be made much more affordable and even made free. This is evident from the word cloud in Fig. \ref{plot_sentiment_4_inkscape}.

The overall sentiment for this query is also positive (Table. \ref{table_sentiment_4_scores}) with a value of $32.75435$ as users are satisfied with the current service. However, they have higher expectation from the policy makers. Around $71\%$ of the users have given positive feedback.  One of the positive statement was \emph{"If government could provide internet in affordable price to everyone it would be great. That would help all kids across the country for study."}. $12\%$ of the people have negative sentiments with one utterance as \emph{"Government should make strict rules related to contents which may negatively harm childrens."}. One comment among the $17\%$ of neutral respondents  was \emph{"I don't know any policies regarding Internet"}. The mean values for positive and negative sentiments are  $0.57196$ and $-0.45413$ respectively. The variances of the same are $0.02156$ and $0.01527$ respectively.
\begin{table}[ht]
  \caption{Sentiment 4 - Scores}
  \centering
  \begin{tabular}{p{2.5cm}|p{1.5cm}|p{3cm}}
  \hline
  \hline
  \textbf{Parameters} & \textbf{Values} & \textbf{Comments}\\  [0.5ex]
  \hline
  Positive & $71$\% & Subjects who expressed positive sentiment\\
  Negative & $12$\% & Subjects who expressed negative sentiment\\
  Neutral & $17$\% & Subjects who expressed neutral sentiment or did not respond to the query\\
  \hline
  Overall & 32.75435 & Adding all the sentiment values\\
  \hline
  Positive(Mean)	&	0.57196 & Mean  of all positive sentiment values\\
  Negative(Mean)	&	-0.45413 & Mean of all positive sentiment values\\
  Positive(Variance)	&	0.02156 & Variance of all positive sentiment values\\
  Negative(Variance)	&	0.01527 & Variance of all positive sentiment values\\
  \hline
  \end{tabular}
  \label{table_sentiment_4_scores}
\end{table}
\subsubsection{Sentiment 5 - What emerging technologies or innovations should the telecom industry consider to improve internet connectivity? Any
infrastructures you wish the citizens had?}
\begin{figure}[ht]
\centering
\includegraphics[width=\columnwidth]{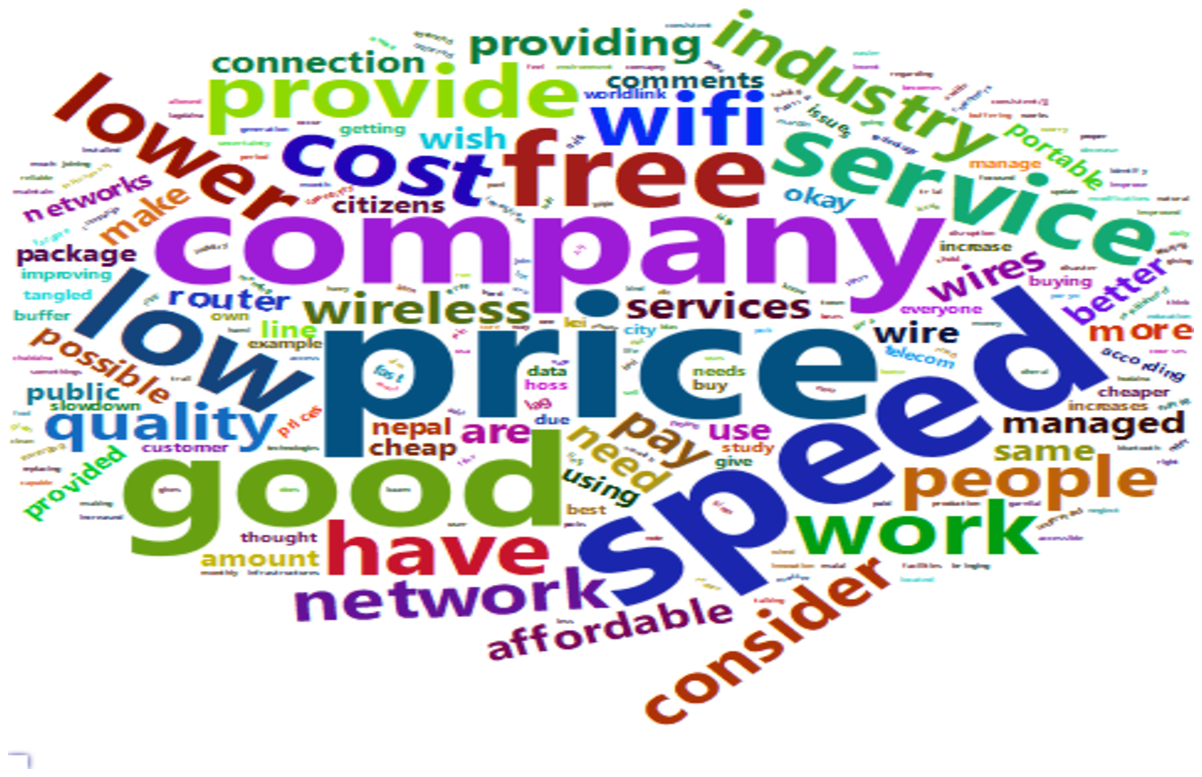}
\caption{Word cloud - Sentiment 5}
\label{plot_sentiment_5_inkscape}
\end{figure}

The response to the above question was also positive. However, the audience was less knowledgable of emerging technologies. Their responses  concentrated on pricing policies and quality of service aspects. They essential expect a good quality, reliable and affordable high internet services. This evident from the word cloud in Fig. \ref{plot_sentiment_5_inkscape}.

The overall sentiment is positive and that of lot of hope for the better with $67$\% of the responses supporting the topic. One of the positive statement \emph{"If there is not any line connection in internet, without line internet then it is more good"}. Only $19$\% are negative with one very dissatisfied comment being \emph{"Company should not fool us by giving best quality Internet during trial period and when we buy the package the Internet is not good. There is a lot of buffering and slowdown."}. Remaining $14\%$ were neutral with one comment being \emph{"I don't know. It don't really matter to me."}. The means of positive and negative sentiments are $0.57873$ and $-0.49813$ respectively. Interestingly, for this question both the variances are close which implies that both sides have similar pattern of opinions.
\begin{table}[ht]
  \caption{Sentiment 5 - Scores}
  \centering
  \begin{tabular}{p{2.5cm}|p{1.5cm}|p{3cm}}
  \hline
  \hline
  \textbf{Parameters} & \textbf{Values} & \textbf{Comments}\\  [0.5ex]
  \hline
  Positive & $67$\% & Subjects who expressed positive sentiment\\
  Negative & $19$\% & Subjects who expressed negative sentiment\\
  Neutral & $14$\% & Subjects who expressed neutral sentiment or did not respond to the query\\
  \hline
  Overall & 27.49387 & Adding all the sentiment values\\
  \hline
  Positive(Mean)	&	0.57873 & Mean  of all positive sentiment values\\
  Negative(Mean)	&	-0.49813 & Mean of all positive sentiment values\\
  Positive(Variance)	&	0.02386 & Variance of all positive sentiment values\\
  Negative(Variance)	&	0.02312 & Variance of all positive sentiment values\\
  \hline
  \end{tabular}
  \label{table_sentiment_5_scores}
\end{table}

\subsubsection{Sentiment 6 - Write your experience of streaming a video youtube channel on their network}
\begin{figure}[ht]
\centering
\includegraphics[width=\columnwidth]{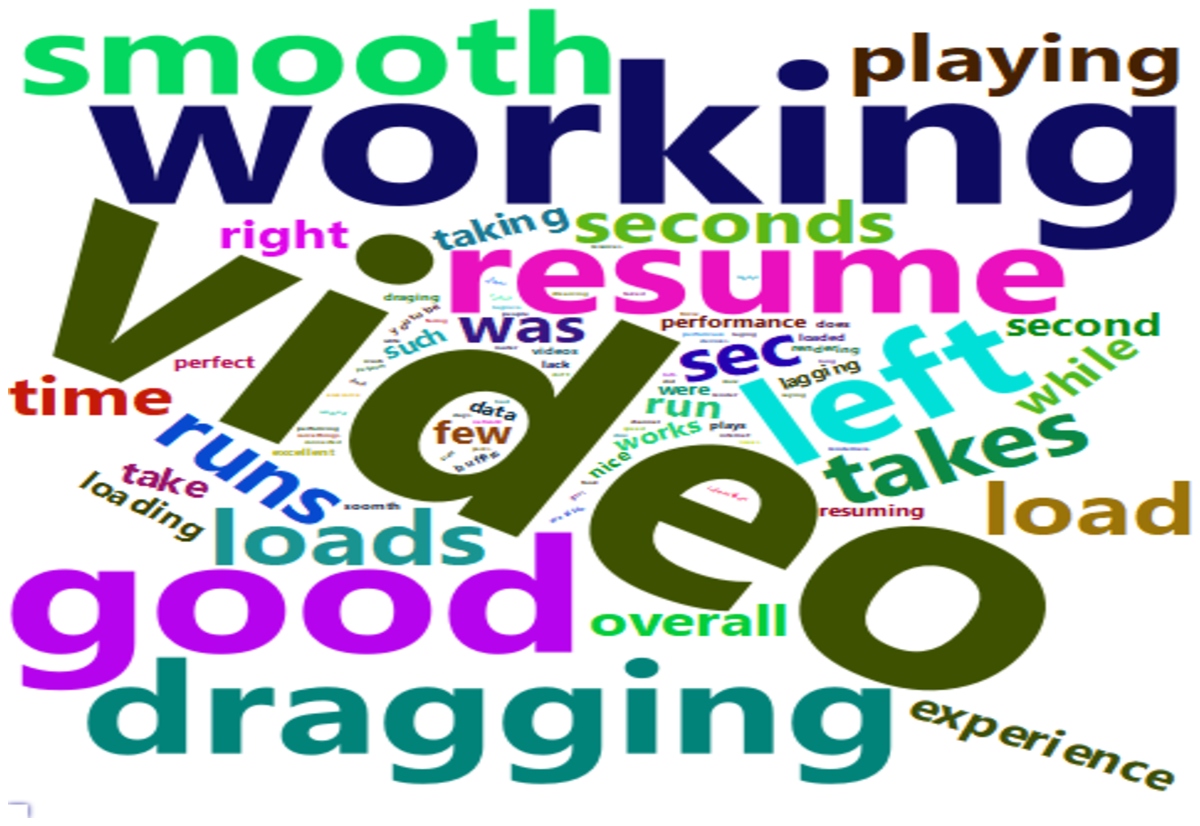}
\caption{Word cloud - Sentiment 6}
\label{plot_sentiment_6_inkscape}
\end{figure}

In this case also, the overall mood is positive. Majority of the respondents say that their youtube experience is good with one comment as \emph{"It is good experience. Video were soomthly playing. Overall, performance was nice"}. Another 10\% finds the experience poor stating \emph{"It was very slow. It takes so much time to load video. It can be also because router is little more far from us. Video were not playing for a long time"}. Remaining 16\% are somewhat satisfied with one impression being \emph{"No buffering runs smooth"}. In this case, the sentiment model seems to give a high score only when the experience is superlative with lots of qualifying words. The word cloud is given in Fig. \ref{plot_sentiment_6_inkscape}.

Again, the overall sentiment is positive $41.32876$ with $74$\% satisfied with the streaming of youtube videos. Around $10$\% of the people are dissatisfied with the service and $16$\% have no issues with their youtube experience.
The mean values for positive and negative sentiments are $0.66766$ and $-0.52665$ respectively. The respective variances are 0.04105 and 0.01202 which shows the negative sentiments are more focussed and pinpointed.
\begin{table}[ht]
  \caption{Sentiment 6 - Scores}
  \centering
  \begin{tabular}{p{2.5cm}|p{1.5cm}|p{3cm}}
  \hline
  \hline
  \textbf{Parameters} & \textbf{Values} & \textbf{Comments}\\  [0.5ex]
  \hline
  Positive & $74$\% & Subjects who expressed positive sentiment\\
  Negative & $10$\% & Subjects who expressed negative sentiment\\
  Neutral & $16$\% & Subjects who expressed neutral sentiment or did not respond to the query\\
  \hline
  Overall & 41.32876 & Adding all the sentiment values\\
  \hline
  Positive(Mean)	&	0.66766 & Mean  of all positive sentiment values\\
  Negative(Mean)	&	-0.52665 & Mean of all positive sentiment values\\
  Positive(Variance)	&	0.04105 & Variance of all positive sentiment values\\
  Negative(Variance)	&	0.01202 & Variance of all positive sentiment values\\
  \hline
  \end{tabular}
  \label{table_sentiment_6_scores}
\end{table}
\subsubsection{Sentiment 7 - Share some of the hopes and aspirations for the future around internet connection and network connectivity. What type of
innovations do you wish to see in 5 years?}
\begin{figure}[ht]
\centering
\includegraphics[width=\columnwidth]{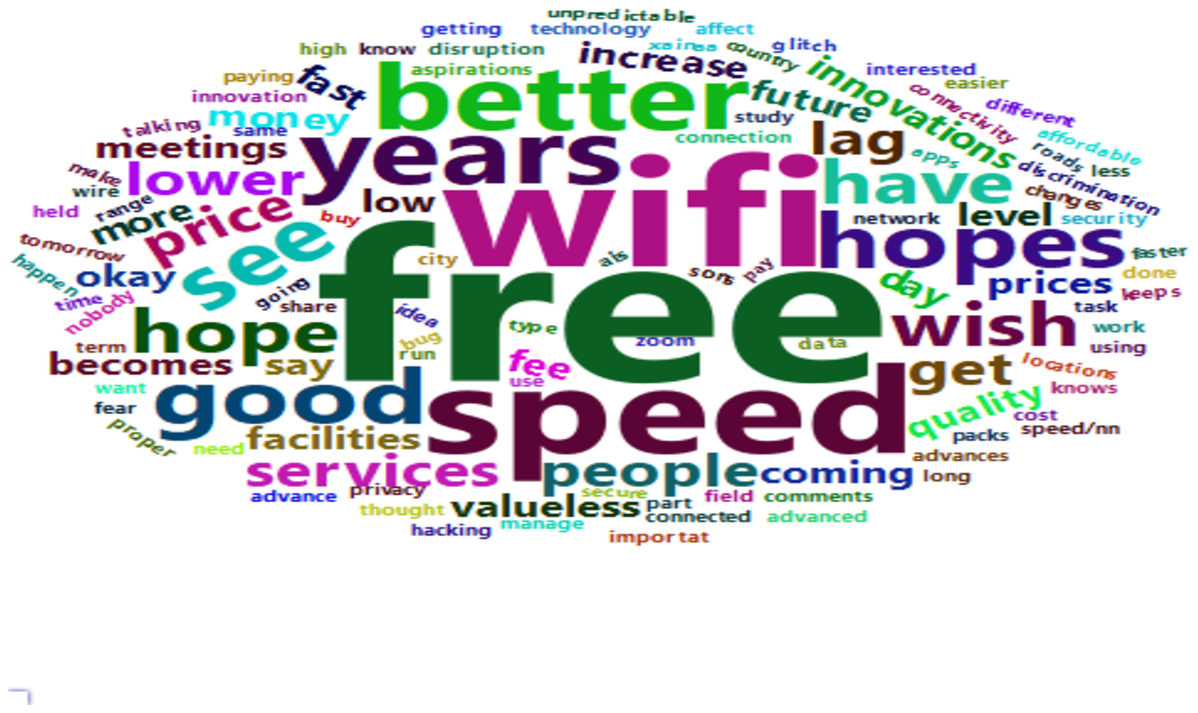}
\caption{Word cloud - Sentiment 7}
\label{plot_sentiment_7_inkscape}
\end{figure}

The response to the equation was mixed with many of the respondents unaware of the value of evolving high speed internet and did not express any opinion. A number of respondents were sceptical about the future. Some of the comments were \emph{"No idea about innovations."}, \emph{"No hopes"} and \emph{"Not thought about it"} which appeared as neutral views. On the positive side one comment with score is \emph{"Free WiFi with good quality Internet"} indicating the price sensitivity of internet services. On the negative side, one comment was \emph{"For now, it is okay how the Internet is and talking about five years nobody knows what will happen tomorrow so I can't really say."}. The word cloud in Fig. \ref{plot_sentiment_7_inkscape} is indicative of these views.

Though the overall sentiment is positive the value of $14.25021$ is lower than some of the previous sentiments. An overwhelming $70$\% views were neutral. $27$\% were positive and a meagre $3$\% had negative views. The mean value of the positive sentiment is 0.62388 and that of negative is -0.44893. Their variances of the same are 0.02666 and 0.00101 respectively which is expected since the number of negative sentiments are much less.
\begin{table}[ht]
  \caption{Sentiment 7 - Scores}
  \centering
  \begin{tabular}{p{2.5cm}|p{1.5cm}|p{3cm}}
  \hline
  \hline
  \textbf{Parameters} & \textbf{Values} & \textbf{Comments}\\  [0.5ex]
  \hline
  Positive & $27$\% & Subjects who expressed positive sentiment\\
  Negative & $3$\% & Subjects who expressed negative sentiment\\
  Neutral & $70$\% & Subjects who expressed neutral sentiment or did not respond to the query\\
  \hline
  Overall & 14.25021 & Adding all the sentiment values\\
  \hline
  Positive(Mean)	&	0.62388 & Mean  of all positive sentiment values\\
  Negative(Mean)	&	-0.44893 & Mean of all positive sentiment values\\
  Positive(Variance)	&	0.02666 & Variance of all positive sentiment values\\
  Negative(Variance)	&	0.00101 & Variance of all positive sentiment values\\
  \hline
  \end{tabular}
  \label{table_sentiment_7_scores}
\end{table}

\subsection{Correlation Matrix}
\begin{figure}[ht]
\centering
\includegraphics[width=\columnwidth]{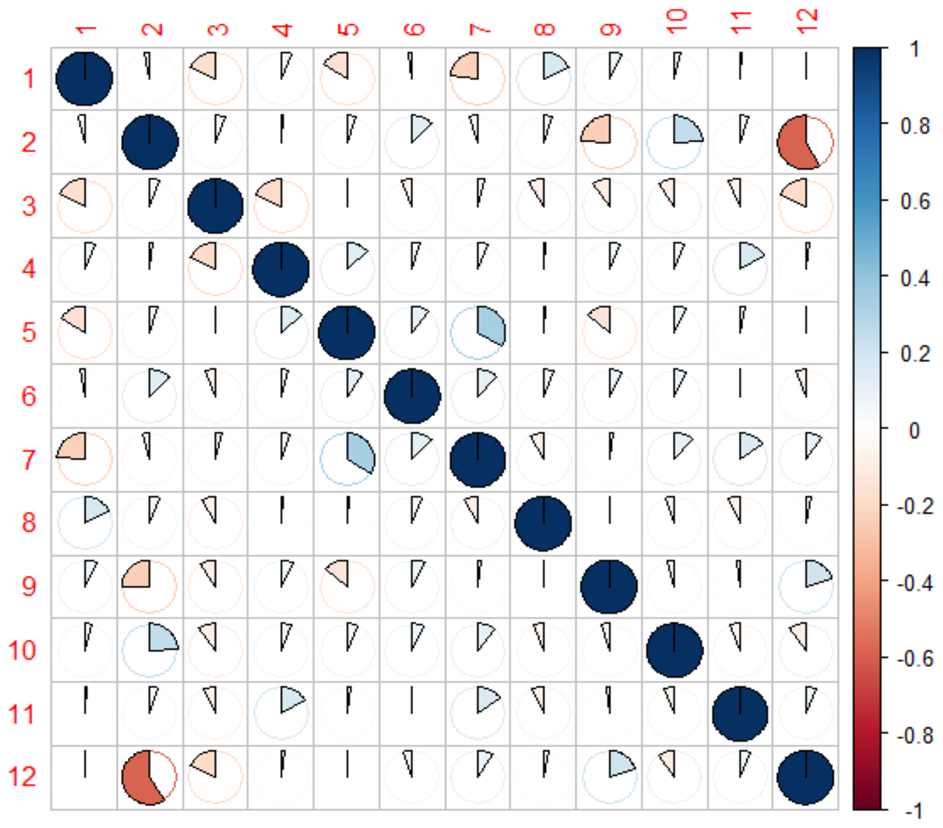}
\caption{Correlation matrix}
\label{plot_correlation_matrix_inkscape}
\end{figure}

Table \ref{table_correlation_matrix_variables} shows the $12$ key  variables of the survey. The correlation matrix of these variables are shown in Fig. \ref{plot_correlation_matrix_inkscape}. The red shades depict the negative correlation and the blues the positive correlation. Thus, the variables 2 and 12 in table \ref{table_correlation_matrix_variables} show a strong negative correlation which means that independent of gender the hopes and aspirations for the future are the same. Similarly, variable 5 and 7 show a strong positive correlation implying the cost of internet is directly related to user rating of data services. This is evident from the prior results also that the expectation of low cost and good quality is high on the agenda of the users.
\begin{table}[ht]
  \caption{Correlation matrix variables}
  \centering
  \begin{tabular}{p{1cm}|p{7cm}}
  \hline
  \hline
  \textbf{Variables} & \textbf{Comments}\\  [0.5ex]
  \hline
  \emph{Q1} & Age of subjects\\
  \emph{Q2} & Gender of subjects\\
  \emph{Q3} & Education of subjects\\
  \emph{Q4} & Occupation of subjects\\
  \emph{Q5} & User rating of data services\\
  \emph{Q6} & How resilient is the network infrastructure to disruptions and disasters? Since 2015 (Earthquake) how has the internet infrastructure changed in your area?\\
  \emph{Q7} & What is the cost of internet services in the area and how does it impact you and your family access to the internet?\\
  \emph{Q8} & How do you report internet connectivity issues or provide feedback about the quality of their services?\\
  \emph{Q9} & Any government policies you wish were in place to better the internet connectivity and infrastructure of Kathmandu?\\
  \emph{Q10} & What emerging technologies or innovations should the telecom industry consider to improve internet connectivity? Any infrastructures you wish the citizens had?\\
  \emph{Q11} & Write your experience of streaming a video youtube channel on their network\\
  \emph{Q12} & Share some of the hopes and aspirations for the future around internet connection and network connectivity. What type of innovations do you wish to see in 5 years?\\
  \hline
  \end{tabular}
  \label{table_correlation_matrix_variables}
\end{table}

\subsection{Principal component analysis}
\begin{figure}[ht]
\centering
\includegraphics[width=\columnwidth]{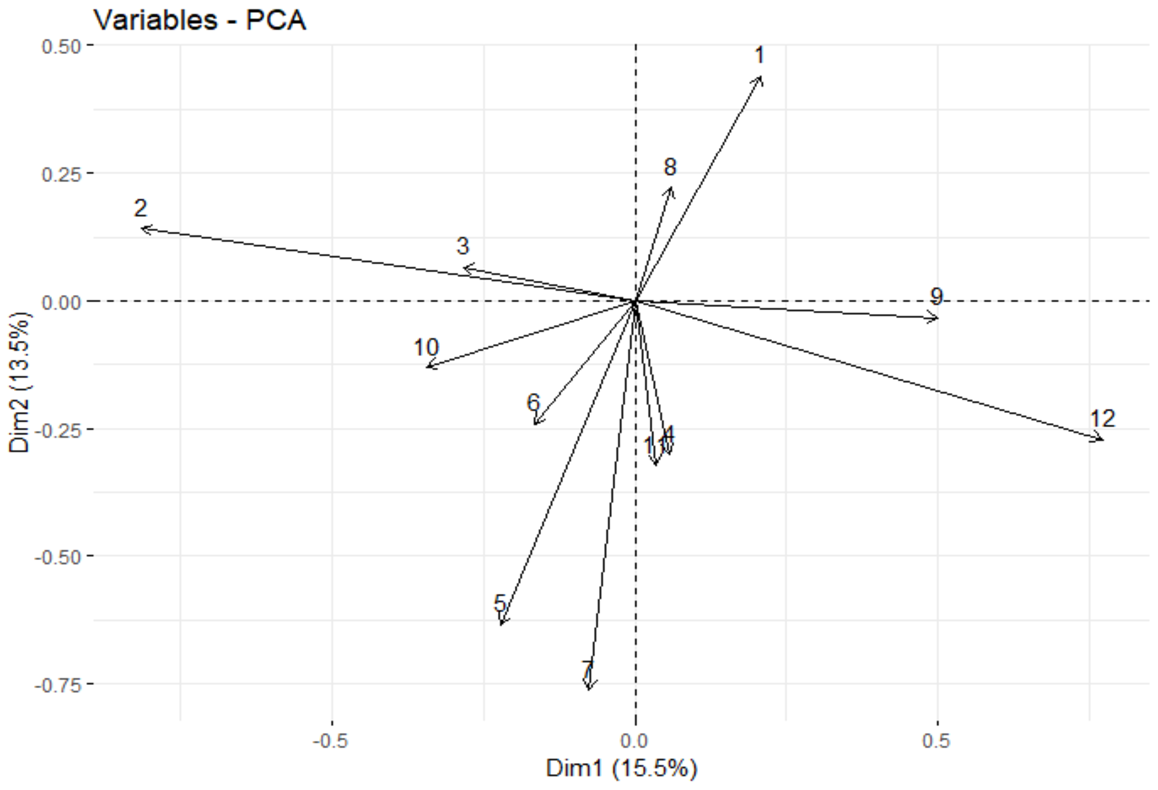}
\caption{PCA vectors of 12 components}
\label{plot_pca_vectors_inkscape}
\end{figure}

\begin{figure}[ht]
\centering
\includegraphics[width=\columnwidth]{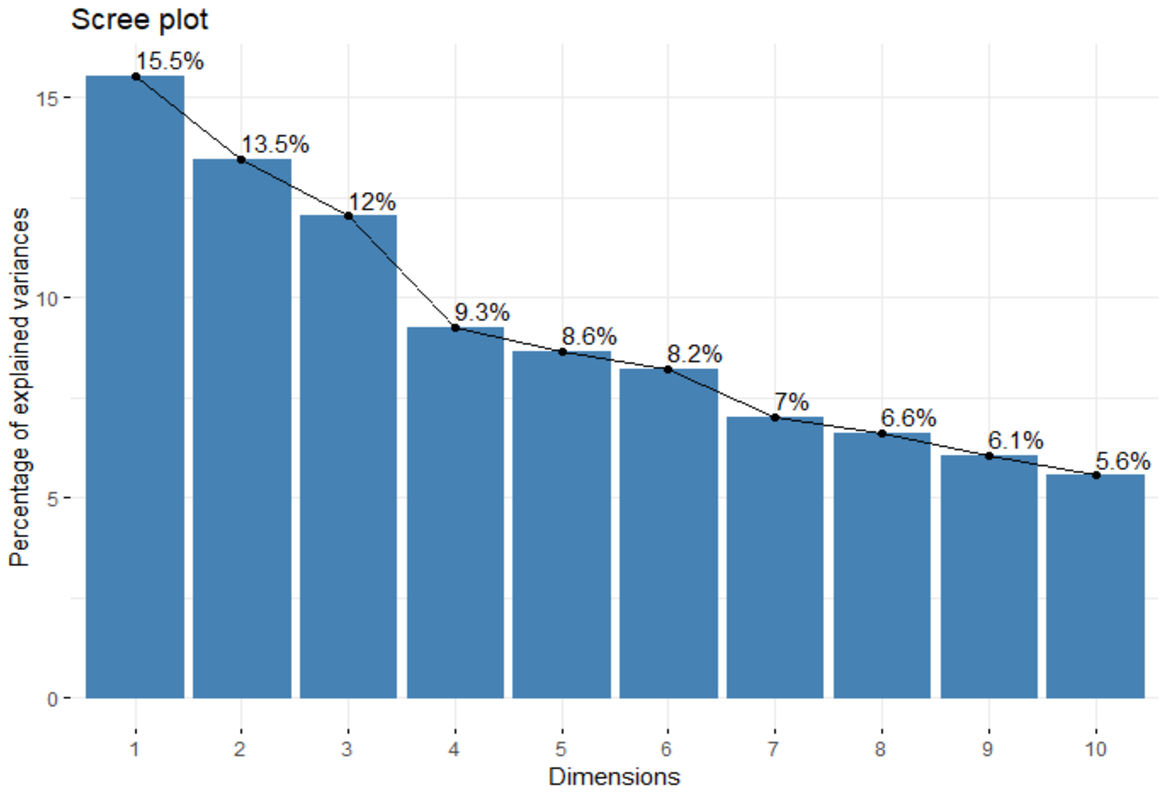}
\caption{PCA plot of top 10 components}
\label{plot_pca_plot_inkscape}
\end{figure}

To find out any dominating factors, PCA is performed on all 12 variables. The 12 vectors are presented in Fig. \ref{plot_pca_vectors_inkscape}. The top 10 components are plotted in Fig. \ref{plot_pca_plot_inkscape}. Both the figures do not show any of the factors to be exceedingly dominant.

\subsection{Clustering}
To identify and clusters in the data \emph{\emph{k-means}} algorithm is applied. First, k-means shows 10 clusters with \emph{betweeness} to \emph{withiness} ratio at 90.9\% on the 12 dimension dataset. The clusters are as follows.
\begin{itemize}
\item \emph{Cluster 1}: Male and with positive sentiment on \emph{Q8}, \emph{Q9} and \emph{Q11} and those who have not responded to question 12.
\item \emph{Cluster 2}: Age group around 40 and satisfied with internet data rate and positive sentiment on \emph{Q8}, \emph{Q11} and \emph{Q12}.
\item \emph{Cluster 3}: Age group around 35 with school level education and  positive opinion on \emph{Q9}.
\item \emph{Cluster 4}: Age group around 30 with positive opinion on \emph{Q9} and \emph{Q10}.
\item \emph{Cluster 5}: Age around 30, satisfied with \emph{Q5}, negative on \emph{Q6} and positive on \emph{Q7} to \emph{Q12}.
\item \emph{Cluster 6}: Age group 25, satisfied with \emph{Q5} and divergent views on \emph{Q6}, very positive on \emph{Q11} and neutral on \emph{Q12}.
\item \emph{Cluster 7}: Age group 45, positive on \emph{Q8}, \emph{Q9} and \emph{Q10 }and neutral on \emph{Q12}.
\item \emph{Cluster 8}: Age below 20, school level education, very satisfied with data rate, negative on \emph{Q6} and positive on \emph{Q7} to \emph{Q11}.
\item \emph{Cluster 9}: Age below 20, below primary education, satisfied with data rate, neutral opinion on \emph{Q6} and positive on \emph{Q8}, \emph{Q9} and \emph{Q10}.
\item \emph{Cluster 10}: Age around 25, below primary education, satisfied with data rate, positive on questions \emph{Q7} to \emph{Q11}.
\end{itemize}
\section{Discussion}\label{section_discussion}
From the above results, some of the broad observations of the survey are as follows. Though the overall satisfaction is high, the bandwidth is far below the recommended 5G limit. This means infrastructure needs to be upgraded. Citizens have high aspirations with high speed, secure and reliable internet services. Internet price sensitivity is a key ingredient which leads to network congestion. Intermittent congestion, meeting high aspirations and exposing people to benefits of high speed internet remains a challenge. Cost effective solution holds the key to resolving the problems.

Sentiment analysis values for the queries are mostly positive. The variances in positive sentiment values are high implying that there is some amount of divergence in the views. However, the variances of negative sentiments are low either because the number of observations are also small or the views are much more pinpointed.  The more the distance between the two means of positive and negative sentiments the more stronger the opposing views. No dominant factors props up with PCA. However, clusters exist in the responses.

\section{Conclusion}\label{section_conclusion}
Internet has made a profound impact on human society from communication, education, health care to business. However, 40\% of the world population do not have access to broadband internet. Though the internet coverage has improved significantly, adoption of the same has been hindered by challenges, such as, affordability, lack of digital literacy, usage experience, and privacy and security concerns, etc. This paper presents an analysis of a field survey internet usage of certain areas of Kathmandu, Nepal, an emerging economy, facing intermittent major network congestion. Three different areas were surveyed asking the subjects about their current experience of internet usage, its impact on their lives and future aspirations. The key finding is that even though the network coverage does not seem to the major issue, the affordability in terms of pricing seems to lead to network congestion. A sentiment analysis is also performed on the inputs received from the survey. The key findings are that the overall sentiment to most queries are positive and future aspirations are high. However, the variance of positive sentiments is high and those for negative sentiment is low. Some correlations are also observed among the queries. No dominant components were observed with PCA though clusters exist in the responses.

Future work would concentrate on providing solutions to congestion providing low cost network deployment, affordable pricing and suggest new business models for providing reliable and cost effective broadband internet to the citizens of emerging economies.


%

\section*{Acknowledgment}
The authors conveys sincere thank Royal Academy of Engineering, United Kingdom for sponsoring this research. The authors would like to thank Utopia Kathmandu, Nepal, for conducting the field survey and University of Bristol for their valuable inputs.

\ifCLASSOPTIONcaptionsoff
  \newpage
\fi



%



\bibliographystyle{elsarticle-num}
\bibliography{IEEEabrv,ref}
%

\end{document}